\documentclass{aa}  

\usepackage{graphicx}
\usepackage{txfonts}
\usepackage{color}

\newcommand{\xmm}{\textit{XMM-Newton}}
\newcommand{\swift}{\textit{Swift}}

\begin{document} 

   \title{Rapid late-time X-ray brightening of the tidal disruption event OGLE16aaa}


   \author{Jari~J.~E. Kajava\inst{1}
          \and
          Margherita Giustini\inst{1}
          \and
          Richard D. Saxton\inst{2}
          \and
          Giovanni Miniutti\inst{1}          
          }

   \institute{Centro de Astrobiolog\'{\i}a (CSIC-INTA), Departamento de Astrof\'{\i}sica, Camino Bajo del Castillo s/n, E-28692 Villanueva de la Ca\~nada, Madrid, Spain\\
    \email{jkajava@cab.inta-csic.es}
    \and
    Telespazio-Vega UK for ESA, Operations Department; European Space Astronomy Centre (ESAC), Camino Bajo del Castillo s/n, E-28692 Villanueva de la Ca\~nada, Madrid, Spain
     \\
    }

   \date{Received April 14, 2020; accepted June 11, 2020}

\abstract{Stars that pass too close to a super-massive black hole may be disrupted by strong tidal forces. OGLE16aaa is one such tidal disruption event (TDE) which rapidly brightened and peaked in the optical/UV bands in early 2016 and subsequently decayed over the rest of the year.
OGLE16aaa was detected in an \xmm\ X-ray observation on June 9, 2016 with a flux slightly below the Swift/XRT upper limits obtained during the optical light curve peak.
Between June 16--21, 2016, Swift/XRT also detected OGLE16aaa and based on the stacked spectrum, we could infer that the X-ray luminosity had jumped up by more than a factor of ten in just one week. No brightening signal was seen in the simultaneous optical/UV data to cause the X-ray luminosity to exceed the optical/UV one.
A further \xmm\ observation on November 30, 2016 showed that almost a year after the optical/UV peak, the X-ray emission was still at an elevated level, while the optical/UV flux decay had already leveled off to values comparable to those of the host galaxy.
In all X-ray observations, the spectra were nicely modeled with a 50--70~eV thermal component with no intrinsic absorption, with a weak X-ray tail seen only in the November 30 \xmm\ observation.
The late-time X-ray behavior of OGLE16aaa strongly resembles the tidal disruption events ASASSN-15oi and AT2019azh.
We were able to pinpoint the time delay between the initial optical TDE onset and the X-ray brightening to $182\pm5$ days, which may possibly represent the timescale between the initial circularization of the disrupted star around the super-massive black hole and the subsequent delayed accretion.
Alternatively, the delayed X-ray brightening could be related to a rapid clearing of a thick envelope that covers the central X-ray engine during the first six months.
} 


   \maketitle
%

\section{Introduction}

Super-massive black holes (SMBH) are found in the centers of galaxies and appear as active galactic nuclei (AGN) when they accrete matter (for a review see, e.g., \citealt{KormendyHo2013,HeckmanBest2014,BrandtAlexander2015}).
In recent years, thanks largely to the dedicated synoptic optical sky surveys, the number of galactic nuclei that have shown unexpectedly rapid optical and X-ray variability has increased dramatically.
One such class of variables are tidal disruption events (TDE), which occur when a star passes so close to the SMBH that it gets ripped apart by strong tidal forces.
A fraction of the disrupted star may end up in bound orbits, leading to strong optical, ultraviolet (UV) and X-ray emission when the fallback material first circularizes around and then falls onto the SMBH \citep{LTH1982,Rees1988,Phinney1989,EvansKochanek1989,StrubbeQuataert2009,GRR2013}.

The basic TDE models predict that after reaching its peak, the brightness evolution of a TDE should follow a power-law decay with an index of $\gamma=-5/3$ \citep{Phinney1989,Rees1990}.
The time scale in which the TDE light curve rises to its peak, as well as the mass fall-back rate, depend on the mass of the SMBH \citep{LKP2009,GRR2013}.
Given that several TDEs follow these predictions rather well, the SMBH masses and the properties of the disrupted stars can be estimated by fitting the light curve \citep{MGRR2019}. 
However, in some cases, the optical/UV decay is not well described by these models.
One pronounced outlier is ASASSN-15lh, which has shown a strong secondary optical/UV brightening \citep{LFS2016}. 
The TDE classification of ASASSN-15lh is not certain, however, as it might be a super-luminous supernova \citep{DSP2016}.  
Another strangely behaving TDE is AT 2018fyk/ASASSN-18ul, which showed a plateau phase after about 50 days into the start of the decline phase \citep{WPvV2019}.
PS16dtm, a likely TDE in a Seyfert I galaxy, also exhibited a plateau phase that lasted for 100 days, while simultaneously having a X-ray luminosity that is a factor of 10 lower than historic levels, suggesting that the AGN was obscured by the TDE debris \citep{BNB2017} or that the flow of material onto the AGN was disrupted by the interaction. 
PS16dtm showed a significant infrared excess during the TDE, perhaps due to reprocessing by a dusty torus in the vicinity of the SMBH \citep{JWY2017}.
Similar complex ``TDE-like'' variations can be seen in general from ``changing-look AGN'' in optical and X-rays (e.g., \citealt{TAM2019}) as well as in the infrared \citep{KKM2017}, which means that care must be taken when identifying an event as a genuine TDE.

Among the X-ray bright TDE candidates, there are a few cases where significant X-ray re-brightening occurs months after the optical/UV peak.
These TDEs have strong thermal X-ray spectral components, with similar temperatures, of about 50--100~eV, as other X-ray detected TDEs (e.g., \citealt{ESF2007,MKM2015,LMI2015,HBA2018}). 
ASASSN-15oi, for example, showed a factor of ten increase of the X-ray flux roughly six months after the optical/UV TDE peak, without any corresponding signal in the optical/UV light curve \citep{GCA2017,HBA2018}.
During this period, the X-ray luminosity significantly exceeded the optical/UV one.
Another TDE that shows puzzling X-ray activity is ASASSN-18jd;
four months into the TDE decay, it showed a bright X-ray flare lasting only a week \citep{NHK2020}.
Similar rapid late-time X-ray flares (or re-brightening events) were reported from AT2019azh \citep{LDS2019,vVGH2020}.
TDEs can also show relativistic jet activity in radio and X-rays, the best example being Swift J1644+57, which has a hard, power-law-like, ``non-thermal'' X-ray spectra \citep{BGM2011, BKG2011, TMG2014}.

This observational diversity is also accompanied by multiple theoretical models for TDEs, which build upon the debris fall-back scenarios \citep{Rees1988,EvansKochanek1989}.
It has been suggested that the initial optical/UV TDE flare is not, in fact, related to accretion onto the SMBH but is, rather, generated by shocks during the initial circularization of the disrupted star around the SMBH \citep{PSK2015, KPS2016}.
An alternative view is that the X-ray source can be shrouded by the stellar debris or by an optically thick envelope that forms during an super-Eddington accretion episode onto the SMBH, and the X-rays are then re-processed into UV/optical light in this envelope \citep{LU1997,StrubbeQuataert2009,MS2016}.
A thick re-processing torus can also be formed around the SMBH if the in-falling gas streams do not cool efficiently \citep{BRL2017}.
As many of these models predict different observational characteristics based on the viewing angle with respect to the newly formed debris disk, meaning an X-ray-bright TDE would correspond to a face-on view of a disk, while an X-ray-faint TDE would be viewed edge-on; thus, a unified model for TDEs has naturally emerged \citep{DMR2018}.

\section{Observations of OGLE16aaa}

\subsection{Discovery and properties of OGLE16aaa}

OGLE16aaa is a TDE candidate, whose discovery was reported by \citet{WLZ2017}. 
It was detected on January 2, 2016 by the OGLE survey at the center of its host galaxy \citep{WKU2016ATel}. 
The PESSTO transient survey took an optical spectrum of it on January 17, 2016 \citep{Firth2016ATel, WLZ2017}, showing a blue continuum with narrow optical emission lines of \ion{H}{$\alpha$}, \ion{H}{$\beta$}, \ion{N}{II,} and \ion{O}{III} at a common redshift of $z=0.1655$ \citep{WLZ2017}.
The optical spectrum resembles the TDE in SDSS J074820.66+471214.6 \citep{YCW2013}, and the line ratios suggested that both weak AGN activity and recent star formation were present in the galaxy \citep{WLZ2017,OCJ2019}.
OGLE16aaa also has rather typical values of rise and decay time scales as well as luminosities for a TDE (for comparison, see, \citealt{vVGH2020}).

\citet{MGRR2019} used the MOSFit TDE models \citep{GNV2018} to derive a black hole mass of $M_{\rm bh} = 3.0^{+1.2}_{-0.8} \times 10^{6}\,M_{\odot}$ for OGLE16aaa. We used this value throughout our study.
We applied the line-of-sight reddening of $E(B-V)=0.028$ \citep{SF2011} and the extinction law of \citet{CCM1989} with the \textsc{redden} \textsc{xspec} model. 
We found the corresponding hydrogen column density of $1.9\times10^{20}$~cm$^{-2}$ using $N_{\rm H} ({\rm cm}^{-2})=(6.86\pm0.27)\times10^{21}~E(B-V)$~mag \citep{GO2009}, and use this value in \textsc{xspec} with the 2016 version of the \textsc{tbabs} model \citep{WAM2000}.
The adopted $N_{\rm H} = 1.9\times10^{20}$~cm$^{-2}$ value is slightly smaller than the one obtained by \citet{HI4PI2016}; $N_{\rm H} \approx 2.7 \times10^{20}$~cm$^{-2}$.
We also assume cosmological parameters derived by \citet{Planck2016}: $\Omega_{\rm m} = 0.308$, $H_0 = 67.8 {\rm~km~s^{-1}~Mpc^{-1}}$, giving a luminosity distance of $D_\textrm{L} = 819.4$~Mpc for a flat Universe.

In this study, we use the previously published, publicly available OGLE \citep{Wyrzykowski2014} V-filter and I-filter light curves of OGLE16aaa.\footnote{http://ogle.astrouw.edu.pl/ogle4/transients/2017a/transients.html} 
The first three months of OGLE photometry were presented in \citet{WLZ2017}, and these data were also utilized more recently by \citet{MGRR2019} and \citet{JNW2019}.
We complement these observations with new \swift/UVOT observations taken in February 2020, as well as archival \xmm\ X-ray and UV observations taken on June 9, 2016 and November 30, 2016.

\begin{figure*}
   \centering
  \includegraphics{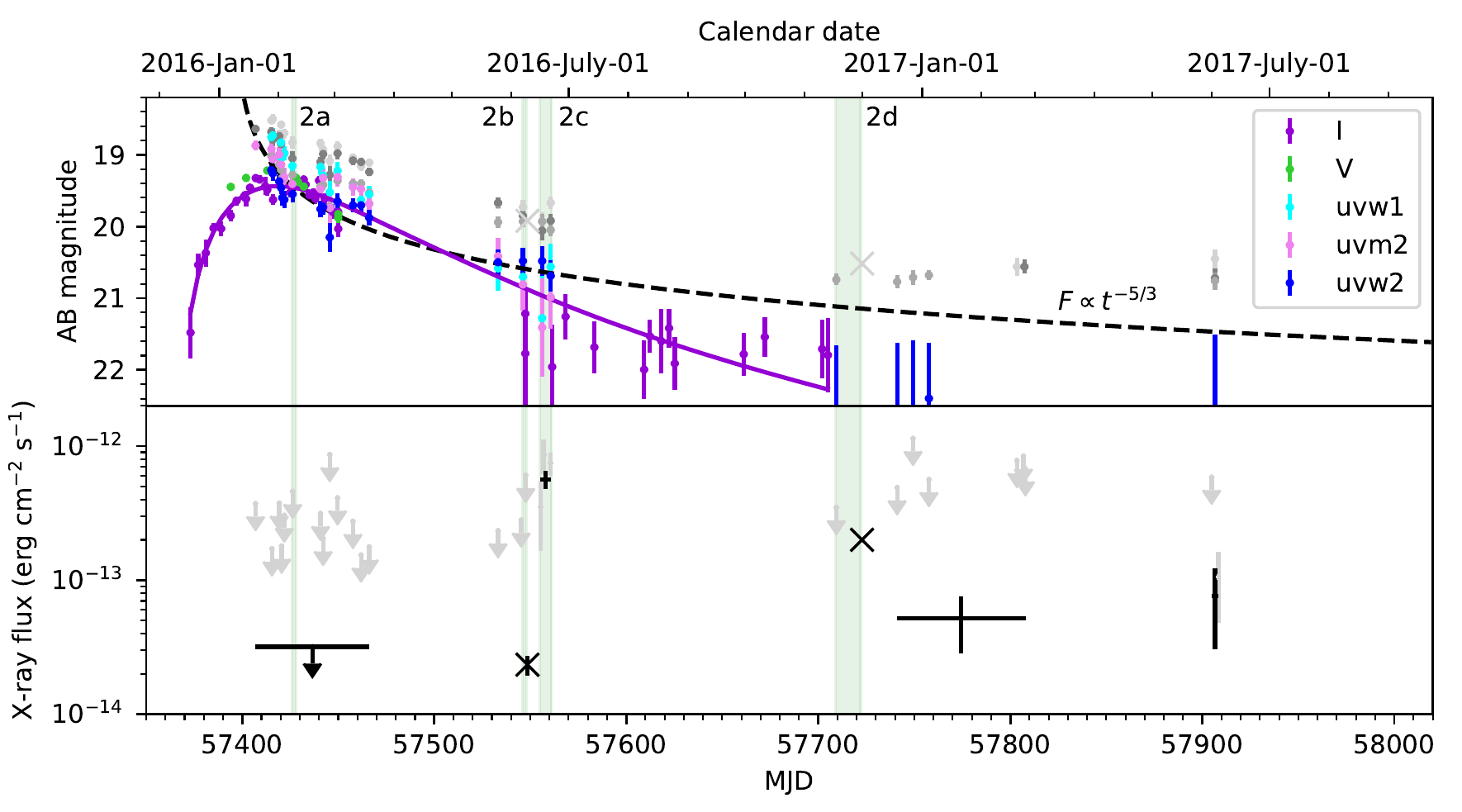}
   \caption{Top panel: Optical and UV light curves of OGLE16aaa. Purple and green data points are the OGLE-I and OGLE-V filter photometry (see \citealt{WLZ2017}), while cyan, violet, and blue are \swift/UVOT uvw1, uvm2 and uvw2 photometry, respectively (the host galaxy contribution has been subtracted). The TDE plus host galaxy AB magnitudes are shown with grey symbols. The purple line is the best-fitting MOSFit TDE model for a $3\times10
  ^6$ solar mass black hole (fitted to the I-filter data, following \citealt{MGRR2019}), and the dashed line shows an arbitrarily scaled $\gamma=-5/3$ powerlaw flux decay trend. Bottom panel: X-ray light curve of OGLE16aaa in the 0.3--1~keV band. The grey arrows denote the upper limits as measured by \swift/XRT during observations of $\sim1$~ks each. The black upper limit comes from stacking 25.7~ks of \swift\ data. During the winter and spring of 2017, XRT detected the OGLE16aaa only marginally. The two black crosses denote the fluxes measured by \xmm. We note the rapid X-ray flux increase between the June 9 \xmm\ observation (MJD 57548) and the June 16--21 stacked \swift/XRT observation (MJD 57555--57560). The time periods for extracting the UV/X-ray spectra shown in Figure \ref{fig:spectra} a), b), c), and d) are shown with green shaded stripes.}
    \label{fig:LC}%
\end{figure*}

\subsection{Swift observations}

OGLE16aaa was observed with the \textit{Neil Gehrels Swift Observatory} (\swift\ hereafter; \citealt{Gehrels2004}) on multiple occasions from January 2016 onward (see \citealt{WLZ2017}, for earlier use of these data near the TDE peak).
In this paper, we re-analyse these archival data, as well as data from our own target-of-opportunity \swift\ observation taken in February 2020, presumably when only the host galaxy was contributing to the optical/UV emission.

The \swift/UVOT photometry was performed by running the \textsc{uvotproduct} tool with a $5\arcsec$ aperture centered on the known position of OGLE16aaa and a $20\arcsec$ sky background region near it. Adjacent measurements were added together until a minimum signal-to-noise ratio (S/N) of 2 was reached, but not allowing observations separated more than 2 days to be co-added.

To obtain the flux densities from the background subtracted and aperture corrected count rates, we used the count rate conversion factors appropriate for a Vega like spectrum (see, \citealt{BBR2016}, their table 1).
This choice is reasonable, given that OGLE16aaa has a relatively featureless thermal continuum with a temperature of about 23000~K \citep{MGRR2019}.
In order to convert the Vega magnitudes obtained from the \textsc{uvotproduct} to AB magnitudes, we followed the calibration document SWIFT-UVOT-CALDB-16-R01.
We then generated \textsc{xspec} compatible pseudo-spectral files and diagonal response matrices for the UVOT filters using the \textsc{flx2xsp} tool.
The UVOT filter central wavelengths and widths were taken from Table 1 of \citet{Poole2008}.

The Swift/XRT X-ray light curve was obtained using the XRT generator online tool \citep{EBP2009}.
We first generated a light curve binned to time resolution of one observation, which is shown in Fig. \ref{fig:LC} using light grey upper limits and error bars when OGLE16aaa was detected.
We also generated a stacked images in the 0.3--1.0 keV band from a few adjacent time intervals. 
For these images, we used the \textsc{ximage} and \textsc{sosta} tools centered on OGLE16aaa coordinates to provide a more stringent upper limit or a source detection.
The XRT generator was also used to obtain the stacked X-ray spectrum between June 16--21, during which OGLE16aaa was detected in individual observations.
A total of 42 X-ray photons were detected in 3379 seconds of stacked exposure during these three observations.
Given the low number of X-ray counts, we grouped the data using \textsc{grppha} to have a minimum of one count per bin and we fitted these data in \textsc{xspec} using Cash fit statistics.

\subsection{XMM-Newton observations}

\xmm\ has observed OGLE16aaa twice; on June 9, 2016 for 15~ks (OBSID: 0790181801) and on November 30 2016 for 36.6~ks (OBSID: 0793183201).
Both observations were done with the EPIC cameras operating in the full frame mode and using the thin optical blocking filter.
The OM observations we performed were carried out using the uvw1 filter and here we utilize the two photometric measurements (AB magnitudes) obtained from the OM pipeline products.

We reduced the EPIC-PN and EPIC-MOS X-ray data using the XMM-SAS version 18.0.0, with the latest calibration files as of January 2020.
First, we generated calibrated event lists using the 
\textsc{epproc} and \textsc{emproc} tools, utilizing the FLAG$==$0 and PATTERN$<=$4 event selection criteria.
We then eliminated soft proton flaring episodes by generating detector averaged light curves above 10~keV and removing periods when the count rates were above 0.7 and 0.35 for PN and MOS, respectively.

OGLE16aaa is clearly detected in images below 1 keV in both observations.
We extracted the light curves and spectra using a circular 20$\arcsec$ source aperture and a 40$\arcsec$ background region from the same EPIC-PN CCD detector. 
The spectra were extracted with the \textsc{especget} tool, which generates the source and background spectra, ancillary response files, and redistribution matrices.
Finally, the spectra were grouped using \textsc{grppha} to have a minimum of 20 counts per bin. 
For the longer observation of November 30, 2016, OGLE16aaa was a factor of ten brighter than in June 9, 2016. 
These EPIC-PN data had sufficient X-ray counts to fit models in \textsc{xspec} using $\chi^2$ fit statistics.
The spectra of PN and MOS detectors matched well in the 0.2--1~keV band and, thus, variable instrument normalizations were not needed in the fits.
In the fainter and shorter observation of June 9, 2016, the MOS detectors saw only a few tens of photons each and, thus, only EPIC-PN data were used.

In the following all errors are quoted at the 1-$\sigma$ level, except for the 3-$\sigma$ upper limits of the \swift/XRT non-detections.

\section{Results}

\begin{table*}
\centering
\caption{\label{tab:xspec}Best-fitting parameters for the June 9 XMM, June 16--21 Swift/XRT, and Nov. 30, 2016 XMM spectra. 
The Galactic absorption column $N_{\rm H}$ was fixed to $1.9 \times 10^{20}$~cm$^{-2}$.
The thermal models are denoted with temperatures $T_{\rm TH}$ (in eV) and normalizations $K_{\rm TH}$ that correspond to a black body ($T_{\rm bb}$, $K_{\rm bb}$), disk black body ($T_{\rm dbb}$, $K_{\rm dbb}$) and bremsstrahlung ($T_{\rm br}$, $K_{\rm br}$). 
The model normalizations are: $K_{\rm bb} = (R_\textrm{bb}$[km]/$d_{10})^2$, $K_{\rm dbb} = (R_\textrm{dbb}$[km]/$d_{10})^2 \cos i$.
The 0.3--1~keV band fluxes are not corrected for local or interstellar absorption and they are given in units of $10^{-13}\,{\rm erg\,cm^{-2}\,s^{-1}}$.
} 
\begin{tabular}{@{}lcccccccr}
\hline\hline
ID & $T_{\rm TH}$ (eV)     & $K_{\rm TH}$             & $\Gamma$                      & $F_{\rm PL~[0.3-1]}$                 & $F_{\rm TH~[0.3-1]}$                         & $F_{\rm TH,~bol}$                  & $\chi^{2}/\textrm{d.o.f.}$ & $\textrm{C}/\textrm{d.o.f.}$\\
\hline
\multicolumn{9}{c}{June 9 XMM-Newton spectrum} \\
\hline
  BB & $52_{-4}^{+4}$  & $2400_{-1400}^{+1700}$ & ...  & ...                & $0.234_{-0.084}^{+0.004}$         & $1.9_{-0.4}^{+0.5}$        & 7.35/7      &  ... \\
 DISKBB & $64_{-5}^{+6}$         & $1000_{-500}^{+900}$ & ...  & ...                & $0.232_{-0.100}^{+0.002}$         & $3.8_{-0.9}^{+1.3}$        & 7.30/7     &  ... \\
\hline
\multicolumn{9}{c}{June 16--21 Swift/XRT spectrum} \\
\hline
  BB & $[52]$  & $58000_{-9000}^{+10000}$ & ...  & ...                & $5.7_{-0.9}^{+0.9}$         & $47_{-7}^{+8}$        & ...      &  10.90/25 \\
 DISKBB & $[64]$         & $25000_{-4000}^{+4000}$ & ...  & ...                & $5.6_{-0.8}^{+0.9}$         & $90_{-20}^{+20}$        & ...     &  10.45/25 \\
\hline
\multicolumn{9}{c}{Nov. 30 XMM-Newton spectrum} \\ 
\hline
  BB & $56.8_{-0.7}^{+0.7}$  & $11300_{-900}^{+1000}$ & ...  & ...                & $2.00_{-0.03}^{+0.03}$         & $12.6_{-0.5}^{+0.5}$        & 138.9/120      &  ... \\
 DISKBB & $69.2_{-0.9}^{+0.9}$         & $5000_{-500}^{+500}$ & ...  & ...                & $1.97_{-0.04}^{+0.02}$         & $24.7_{-1.0}^{+1.1}$        & 120.3/120     &  ... \\
  BREMS & $103_{-2}^{+2}$         & $7.7_{-0.4}^{+0.5}  \times 10^{-3}$ & ...  & ...                & $1.92_{-0.03}^{+0.03}$         & ...        & 114.5/119     &  ... \\
 DISKBB+PL & $67.0_{-1.1}^{+1.1}$         & $6100_{-600}^{+700}$ & $[1.8]$  & $0.054_{-0.015}^{+0.015}$                & $1.91_{-0.03}^{+0.03}$         & $26.6_{-1.2}^{+1.3}$        & 109.3/119     &  ... \\

\hline 
\end{tabular}
\end{table*}

The optical/UV and the X-ray light curves of OGLE16aaa are shown in Fig. \ref{fig:LC}.
\citet{WLZ2017} was able to pinpoint the first TDE detection to MJD~$57369\pm4$ (Dec. 13, 2015).
As noted by \citet{JNW2019}, the I-filter light curve has two peaks, with the maximum reached on January 20, 2016 (MJD 57403), giving a TDE rise time of $\approx34$~days.

Based on the February 9, 2020 \swift/UVOT observations, we measured the following AB magnitudes: $20.93\pm 0.18$, $20.43\pm  0.21$, $20.30\pm 0.19$~mag, for the uvw2, uwm2, and uvw1 filters, respectively. 
We also clearly detected the host galaxy of OGLE16aaa in the u and b filters and used the color correction terms (from the SWIFT-UVOT-CALDB-03-R02 calibration document) to derive the Johnson B magnitude of $B = 18.91 \pm 0.14$~mag.
This magnitude is, in fact, slightly higher than the USNO-B1.0 catalog value of $B = 18.38$~mag (with a typical uncertainty about 0.3~mag), suggesting that we were indeed measuring only the host galaxy contribution.
We therefore subtracted these Feb. 9, 2020 host galaxy magnitudes from all photometry taken in 2016--2017 (and propagated the errors), such that only the TDE contribution is shown for the UV filters in Fig. \ref{fig:LC} (similarly to the host-galaxy subtracted OGLE I and V filter data).

The \swift/UVOT magnitudes are consistent with the host galaxy from about MJD 57700 (November 2016) onward, apart from the uvw2 filter data, which is slightly above the host galaxy level. 
The host galaxy-subtracted \swift/UVOT AB magnitudes of OGLE16aaa are shown in cyan, pink, and blue points in Fig. \ref{fig:LC}.
The non-subtracted photometry is shown with grey points, and we note here that the \swift/UVOT AB magnitudes are higher than in \citet{WLZ2017} by factors consistent with the Vega-AB magnitude conversion, suggesting that \citet{WLZ2017} reported the UVOT magnitudes in the Vega system.
The optical/UV light curve indicates a flux decay that is roughly consistent with the standard $\gamma = -5/3$ trend, particularly in the uvw2 filter.
The I-filter brightness evolves slightly more rapidly, having fallen off to a barely detectable level by early June 2016, and it is much better described by the MOSFit TDE models with $M_{\rm bh} = 3.0 \times 10^{6}\,M_{\odot}$ (purple line in Fig. \ref{fig:LC}; see also \citealt{MGRR2019}).
However, around the same time, the \swift/UVOT magnitudes are still significantly higher compared to the host galaxy values.

The X-ray light curve of OGLE16aaa is shown in the bottom panel of Fig. \ref{fig:LC}.
\xmm\ detected it in both observations (crosses in Fig. \ref{fig:LC}).
On June 9, 2016, the 0.3-1~keV black body flux was about $2.3\times10^{-14}~\textrm{erg}~\textrm{cm}^{-2}~\textrm{s}^{-1}$, while on November 30, 2016, the flux was ten times higher at $F_{\rm bb~[0.3-1]~keV}\approx 2.0\times10^{-13}~\textrm{erg}~\textrm{cm}^{-2}~\textrm{s}^{-1}$ (see Table \ref{tab:xspec}).
\swift/XRT, on the other hand, did not detect OGLE16aaa in the first three months: not during individual observations (see the gray upper limits), nor by stacking all the 13 observations, with a 3-$\sigma$ upper limit of $6.93\times10^{-4}$ count~s$^{-1}$ (or $F_{\rm bb~[0.3-1]~keV} \lesssim 3.18\times10^{-14}~\textrm{erg}~\textrm{cm}^{-2}~\textrm{s}^{-1}$, see the black upper limit in Fig. \ref{fig:LC}).
The measured XRT count rates (and upper limits) were converted to 0.3--1~keV fluxes using a conversion factor of $4.589\times10^{-11}$, which was derived using the measured 0.3--1~keV count rate and the observed flux in the stacked June 16--21 \swift/XRT spectrum.
Interestingly, the first \swift/XRT detection came only a week after the first \xmm\ observation, with a flux more than ten-fold higher.
That is, in just a matter of one week, OGLE16aaa brightened by a factor of more than ten in X-rays, without any hints of a simultaneous brightening in the UV nor optical light curves.
This allows us to constrain the date of the X-ray brightening to MJD 57551$\pm3$, taking place 182$\pm5$ days after the onset of the optical TDE. 

\begin{figure*}
  \begin{tabular}{cc}
    \includegraphics[width=0.48\textwidth]{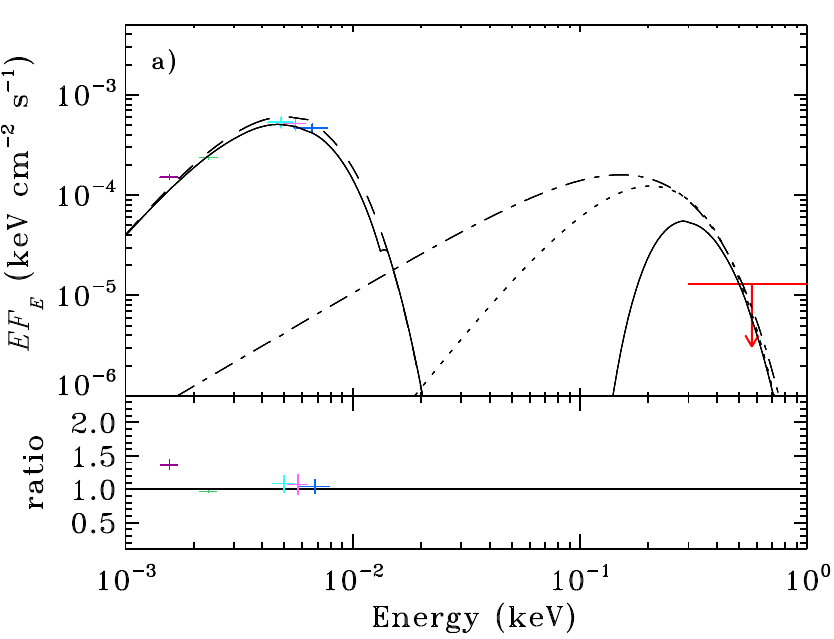} &  
    \includegraphics[width=0.48\textwidth]{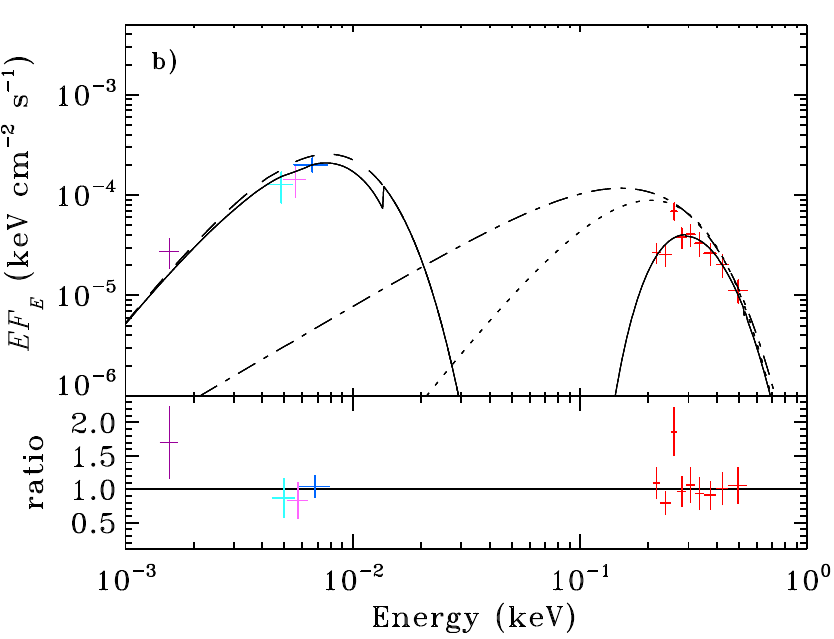} \\
    \includegraphics[width=0.48\textwidth]{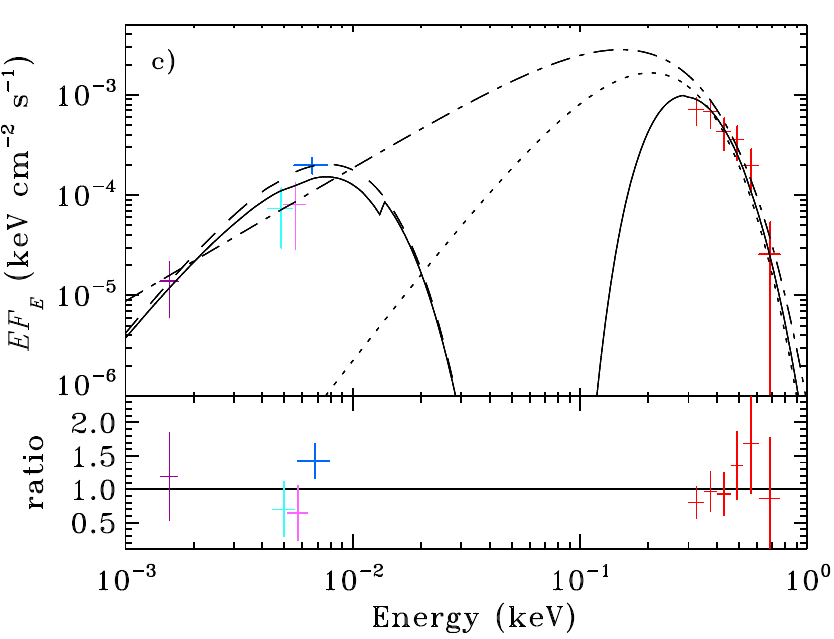} &
     \includegraphics[width=0.48\textwidth]{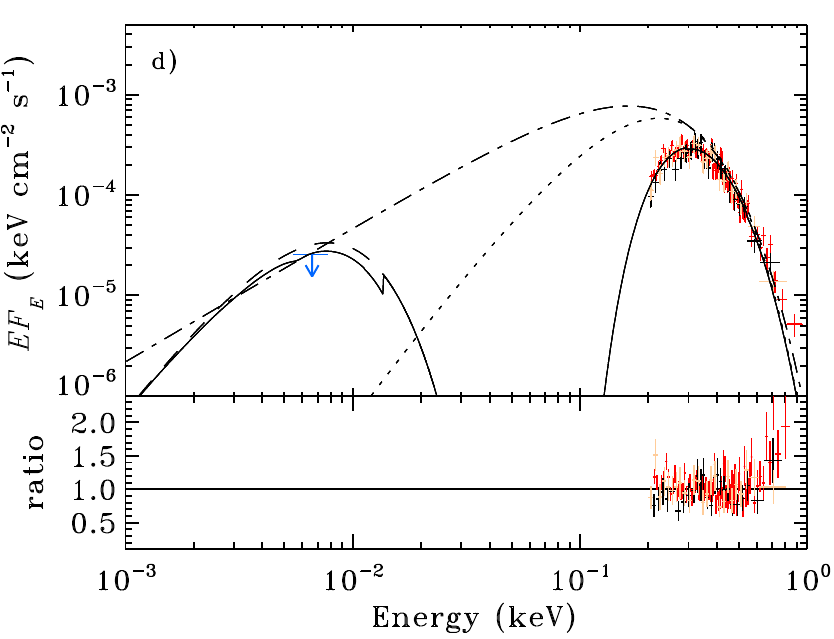}
  \end{tabular}
  \caption{Absorbed dual black body spectral fits of OGLE16aaa during a) the Spring 2016 optical/UV peak, b) June 9, 2016 \xmm\ observation, c) during the stacked June 16-21 Swift/XRT observation and d) the November 30, 2016 \xmm\ observation. The OGLE I and V filter data are shown with purple and green points, the \swift/UVOT uvw1, uvm2, and uvw2 data with cyan, pink, and blue, the \xmm/PN and \swift/XRT data are shown in red, while \xmm/MOS1 and MOS2 are shown in black and brown, respectively. The unabsorbed, and de-reddened black bodies corrected for Milky Way absorption are shown with dotted and dashed lines, while the unabsorbed \textsc{diskbb} component is shown with a dot-dashed lines. We note the large positive residuals in panel d), indicating a presence of a weak, but significant, high-energy X-ray tail.}
\label{fig:spectra}%
\end{figure*}

The optical/UV/X-ray spectral energy distributions of the spring 2016 \swift/XRT non-detection, the June 9, 2016 \xmm\ observation, the stacked June 16--21 \swift/XRT observations, and the November 30, 2016 \xmm\ observation are shown in panels a), b), c), and d) of Fig. \ref{fig:spectra}. 
These four epochs are also highlighted in Fig. \ref{fig:LC} with green stripes.
The two marginal XRT detections in 2017 are based on five and three X-ray photons, and are, thus, not suitable for spectral analysis.
The four X-ray spectra can be adequately fitted with a cool 50--70~eV thermal model in the X-rays, the peak of the emission always below the X-ray band (see Table \ref{tab:xspec}).
There is no need to add an absorber apart from the interstellar absorption column of $N_{\rm H,~gal} = 1.9\times10^{20}$~cm$^{-2}$.
The optical/UV data is also consistent with a thermal component in the Rayleigh-Jeans regime that peaks just above the UV band, that is, reddened according to the interstellar value of $E(B-V)=0.028$ \citep{SF2011}.

During the spring 2016 optical/UV peak, the black body temperature and radius are $T_{\rm bb,~cold} = 14800\pm500~K$ and $R_{\rm bb,~cold} = (1.77\pm0.07)\times10^{15}$~cm (assuming a luminosity distance of $D_{\rm L} = 819.6$~Mpc), although the discrepant I-filter data point makes this (and all other) fits statistically unacceptable.
No X-ray emission is detected from OGLE16aaa and we can derive a limit on the ratio between the X-ray emitting hot black body luminosity and the colder, $\sim20000$ K optical black body luminosity $L_{\rm bb,~hot}/L_{\rm bb,~cold} \lesssim 0.20$ during the optical peak.

On June 9, 2016 the optical/UV data is consistent with the reported $T_{\rm bb} \approx 23000$~K \citep{WLZ2017,MGRR2019}, and the corresponding emission radius is $R_{\rm bb} = (4.8\pm0.3)\times10^{14}$~cm.
The temperature evolution is consistent with the clear difference of UV colors between the early 2016 TDE UV peak compared to June 9, 2016.
The X-ray black body component is still weaker than the UV one, the luminosity ratio now being $L_{\rm bb,~hot}/L_{\rm bb,~cold} \approx 0.35$.
The X-ray spectrum is well described with a $T_{\rm bb,~hot} \approx 50$~eV, but it can also be modeled with the \textsc{diskbb} model \citep{MIK1984} or with a thermal Bremsstrahlung model with roughly a factor of two higher temperature compared to the black body fits.
When extrapolating the best fitting disk or Bremsstrahlung models down to the UV and near infrared (NIR) measurements by \swift/UVOT and OGLE-I, we find that the measured UV and NIR fluxes are factor of about 40 or 10 higher than predicted by the model, respectively.
This allows us to exclude the possibility that the entire NIR to X-ray spectrum is generated by a single temperature optically thin Bremsstrahlung component or by a standard multi-color disk during this epoch.

In the stacked June 16--21 \swift/XRT spectrum shown in panel c) of Fig.~\ref{fig:spectra}, we cannot constrain the temperature well and so we fix it to the value measured by \xmm/PN the week prior.
The obtained black body radius corresponds to $R_{\rm bb} \approx 1.97 \times 10^{12}$~cm (assuming a luminosity distance of $D_{\rm L} = 819.6$~Mpc). 
This is slightly larger than the Schwarzschild radius for a $3\times10^{6}$ solar mass black hole of OGLE16aaa \citep{MGRR2019}. 
The corresponding bolometric luminosity estimate of this X-ray hot thermal component is $L_{\rm bb,~bol} = (3.8\pm0.6)\times 10^{44}~\textrm{erg}~\textrm{s}^{-1}$, which is exactly the Eddington limit for a $3\times10^{6}$ solar mass black hole.
For the \textsc{diskbb} model the estimated bolometric luminosity is roughly two times higher.
The X-ray black body now dominates the bolometric luminosity, with the ratio being $L_{\rm bb,~hot}/L_{\rm bb,~cold} \approx 8.14$.

In the November 30 \xmm\ observation, the disk model \textsc{diskbb} fits the X-ray spectrum better than a simple absorbed black body ($\Delta\chi^2=18.6$ for the same number of degrees of freedom (DOF)).
The best fitting temperatures are consistent with the June 9 \xmm\ observation, despite the variation by a factor of ten of the observed flux. 
However, even the \textsc{diskbb} model leaves an excess near 1~keV, as shown in the bottom panel of Fig. \ref{fig:spectra}d.
This excess can be modeled away using a $\Gamma=1.8$ power law (the slope cannot be constrained), which carries about 3\% of the 0.3--1~keV flux and becomes the dominant spectral component above 0.8~keV.
The ratio between the X-ray and optical black body luminosities had further increased to $L_{\rm bb,~hot}/L_{\rm bb,~cold} \gtrsim 17.4$.
Curiously, in the November 30 \xmm\ observation and in the June 16-21 \swift\ observation, the extrapolated \textsc{diskbb} component (marked with a dot-dashed lines in Fig. \ref{fig:spectra}) fitted only to the X-ray data matches very well the quasi-simultaneous optical/UV fluxes.
It is therefore not obvious if at later times we are seeing two distinct thermally emitting regions.
In fact, if we take the best-fitting \textsc{diskbb} normalization from November 30 and fix the June 16-21 \swift/XRT \textsc{diskbb} normalization to this value (thus by definition $L \propto T^4$), we still obtain a good fit to the June 16-21 \swift/XRT data with a temperature of $T_{\rm in} = 71 \pm 2$ eV and a $\textrm{C-stat} = 12.17$ for 25 DOF (i.e., $\Delta \textrm{C-stat} = 1.72$ compared to the best fitting value shown in Table \ref{tab:xspec}).
This demonstrates that with the available data, we cannot reject a scenario where a standard disk forms and reaches the ISCO on June 16th, and cools thereafter according to the $L \propto T^4$ relation. 

The November 30 \xmm\ observation can be used to derive two useful constraints on the local absorption.
On the one hand, if in addition to the interstellar absorption by the Milky Way we add a cold absorber in the vicinity of OGLE16aaa, we can find a 3-$\sigma$ constraint of $N_{\rm H} \lesssim 0.7\times10^{20}$~cm$^{-2}$ from this high S/N \xmm\ spectrum.
On the other hand, we can also check how much additional local absorption would be required to only attribute the differences between the June 9 and the November 30 \xmm\ observations to changes in local absorption.
We thus fixed the X-ray spectral parameters to values obtained from the November 30 \xmm\ spectrum, and tested three cases with the June 9 \xmm\ data.
First, a neutral fully covering absorber can be ruled out by the data, since we obtain $N_{\rm H} \approx 2.4\times10^{21}$~cm$^{-2}$ in a very poor fit with $\chi^2 = 77.1$ (for 8 DoF).
Second, we added a neutral partial covering absorber with \textsc{tbpcf} to the best fitting November 30 \xmm\ model, which fits the June 9 data much better ($\chi^2 = 8.5$ for 7 DOF). 
We obtained a partial covering fraction of $0.88 \pm 0.01$, with a column that should be higher than $N_{\rm H} \gtrsim 7.3\times10^{21}$~cm$^{-2}$.
Thirdly, we added the ionized absorber model \textsc{zxipcf} and found that for a fully covering column of $N_{\rm H} = 1.4^{+0.6}_{-0.5} \times10^{22}$~cm$^{-2}$ and an ionization parameter of $\log \xi = 0.47^{+0.09}_{-0.12}$, the data can also be fitted well ($\chi^2 = 7.3$ for 7 DOF).
Therefore, in order to attribute the X-ray non-detection in Spring 2016 to absorption, a higher column or covering factor, or a lower ionization state than quoted above, would be required.

\section{Discussion}

The observed late-time X-ray behavior of OGLE16aaa strongly resembles that of the TDEs ASASSN-15oi \citep{GCA2017,HBA2018} and AT2019azh \citep{LDS2019,vVGH2020}.
ASASSN-15oi was monitored relatively consistently for the first 100 days since its discovery, after which there was a 100 day gap in the UV and X-ray coverage.
Initially the UV flux was factor 100-1000 higher than the X-ray flux.
When the regular X-ray and UV monitoring was resumed, the largest changes in the UV versus X-ray flux ratio had already occurred, but there was still a gradual X-ray flux increase for another 100 days or so (see \citealt{GCA2017}, their Figs. 3 and 4).
Very similarly, in AT2019azh the early coverage lasted about 110 days since discovery and when the UV and X-ray monitoring was resumed about 120 days later, the X-ray flux had increased by more than factor of ten during this long gap. 

In the case of OGLE16aaa the X-ray and UV coverage by \xmm\ and \swift\ in June was very fortunate, allowing us to pinpoint the X-ray brightening to within a seven-day interval between June 9 and June 16, 2016, taking place 182$\pm$5 days from the onset of the optical TDE.
The measured delay between the optical versus X-ray brightening is consistent among the three TDEs, but it is only for OGLE16aaa that the X-ray brightening can be determined to be very abrupt.
If the X-ray and optical emission is produced by two distinct thermal components, the bolometric luminosity of the hotter X-ray black body was at least a factor of 8--17 times higher than the optical black body luminosity, being roughly comparable to ASASSN-15oi, AT2019azh as well as ASASSN-14li (see \citealt{LDS2019}; their Fig. 10) after considering that in previous works, the X-ray luminosity was estimated in the 0.3-10 keV band, which only carries about 12--16 per cent of the bolometric flux (see Table \ref{tab:xspec}).

The similarities among the three TDEs are not limited to the light curve morphology.
All three TDEs had very similar X-ray black body temperatures of 40--70 eV.
Also, in all three of them, the black body temperature remains constant while the X-ray flux increases by more than factor of ten.
All three TDEs show signs of a high energy X-ray tail, which in the case of OGLE16aaa is marginally detected and carries only 3\% of the 0.3--1 keV flux.
Moreover, the phenomenological X-ray spectral analysis of OGLE16aaa optical/UV and X-ray data shows a series of interesting coincidences.
After the X-ray brightening event that occurred between June 9 and June 16, the hot thermal component that contributes to the X-rays has a luminosity of $L_{\rm bb,~bol} = (3.8\pm0.6)\times 10^{44}~\textrm{erg}~\textrm{s}^{-1}$, which matches the Eddington value within errors for the $3\times10^{6}$ solar mass black hole of OGLE16aaa \citep{MGRR2019}.
ASASSN-15oi had a similarly high X-ray luminosity after its brightening event, while in AT2019azh, the luminosity was an order of magnitude lower.
When extrapolating the \textsc{diskbb} component down to the NIR range after the X-ray brightening, the optical/UV spectral energy distribution of OGLE16aaa is fully consistent with a disc spectrum and, furthermore, the corresponding black body radius is of the order of the Schwarzschild radius, just as in ASASSN-15oi.
The viscous time scale in the disk is \citep{Pringle1981} $t_{\rm visc} =  R^{3/2}\alpha^{-1}(H/R)^{-2}(GM)^{-1/2} \approx 5.4$~days (assuming $\alpha=0.1$, $H/R=0.1$, $R = 10 R_{\rm g} = 4.45 \times 10^{12}$~cm, $M=3\times10^6\,M_{\odot}$) and so the observed X-ray flux increase time scale of less than one week would be compatible to the viscous time scale of the innermost X-ray emitting region of the accretion flow.
However, we stress that if both the UV and X-ray emission originates from a standard disk, the X-ray brightening cannot be associated with a global increase in mass accretion rate in the whole disk, otherwise we should have seen a leading UV brightening event prior to the X-ray brightening on June 16th. 

From the sparse X-ray data the X-ray flux in OGLE16aaa appears to decay smoothly after the brightening event of June 16–21, 2016 although we cannot exclude that the brightening event was a short lived flare similar to the ones seen in
ASASSN-18jd \citep{NHK2020} or AT2019ehz/Gaia19bpt \citep{vVGH2020}, or -- perhaps more exotically -- if they are related to X-ray quasi-periodic eruptions recently detected in two super-soft galactic nuclei GSN 069 \citep{MSG2019} and RX J1301.9+2747 \citep{GMS2020}.
In the latter two sources, the thermal component during the flare is significantly hotter, however, namely, it is more in line with the $T_{\rm bb} \sim 150$ eV soft excess component seen in several AGN \citep{GD2004,DDJ2012}.

The X-ray temperatures are consistent with respect to their being constant between the two \xmm\ observations where they can be confidently measured, despite the factor of ten increase in flux.
This is not typically expected for a disc that extends to the innermost stable circular orbit (ISCO).
In fact, in the June 9 and November 30 \xmm\ observations, the black body radii are smaller than the ISCO for a non-rotating Schwarzschild black hole.
This casts a doubt on the question of whether the X-ray brightening is really due to a sudden increase of the accretion rate at the innermost regions of the accretion disc around the SMBH.
However, for accretion disks near or above the Eddington limit, the innermost parts of the flow may not reach thermodynamical equilibrium and can ``overheat'' due to the low density \citep{Beloborodov1998}.
This non-standard inner flow can manifest itself as a variable spectral hardening factor \citep{Davis2019} that could cause the counter-intuitive thermal evolution we observe.

There are various ways to interpret the late-time X-ray brightening seen in OGLE16aaa. 
In the \citet{PSK2015} framework (see also \citealt{SKC2015,KPS2016}), the initial optical/UV flare would be generated by shocks during the circularization process of the disrupted star around the SMBH, while the late-time X-ray brightening would arise from a delayed accretion of this circularized gas onto the SMBH.
\citet{GCA2017} and \citet{LDS2019} favored this model for explaining the delayed X-ray emission in ASASSN-15oi and AT2019azh.
In the context of this model, OGLE16aaa presents the best case study so far, thanks to the precise measurement of the relevant time scales; the optical TDE onset is known to occur in MJD~$57369\pm4$, optical TDE rise time is about 34 days, the X-ray brightening happens during a 7-day interval, and, thus, the X-ray brightening delay from the optical TDE onset is known to be $182\pm5$ days. 
However, it is not clear why OGLE16aaa as well as ASASSN-15oi and AT2019azh have such long delays between the peak optical and X-ray brightness, when in other ``textbook TDEs,"\ such as ASASSN-14li \citep{vVAS2016,HKP2016,KPS2016,MummeryBalbus2020} and XMMSL1 J0740-85 \citep{SRK2017}, the brightening occurs simultaneously.
OGLE16aaa has one of the largest inferred black body radii of $R_{\rm bb} \approx 1.8\times10^{15}$~cm during the optical TDE peak (for comparison, see, \citealt{vVGH2020}), which could indicate a larger than usual circularization radius that would, thus, possibly be the cause of the 182-day delay of the Eddington limited accretion onto the SMBH.
The fact that at late times, both the optical and X-ray data are consistent with being drawn from only one spectral component lends support to the formation of a standard accretion disc around the SMBH at this time.

Another scenario for the late-time X-ray brightening could be related to variable local absorption.
Several TDEs show enhanced X-ray absorption with respect to the expected Galactic absorption column \citep{AGRR2017}.
This excess absorber could be the TDE debris and the recently formed complex accretion flow onto the SMBH or it may just be due to the interstellar gas and dust in the host galaxy.
In OGLE16aaa, there are no clear signs of this excess local absorption; all X-ray spectra can be well-fitted with thermal models that have a column fixed to the Galactic value.
The absorber could, however, be patchy or ionized and, indeed, we cannot reject a model in which the difference between the two \xmm\ spectra taken before and after the X-ray brightening is simply attributed to a partial covering or an ionized local absorber covering the X-ray source prior to the brightening.
Therefore, the OGLE16aaa data allow us to put constraints to the TDE models which assume that the X-ray emission is re-processed into optical/UV light in a thick torus or a wind or outflow that covers the X-ray-emitting inner accretion flow near the SMBH (e.g.,  \citealt{LU1997,StrubbeQuataert2009,MS2016,LuBonnerot2020}).
The only viable sequence of events in the context of these models would be that from the TDE onset until the first June 9 \xmm\ observation, the obscuring material near OGLE16aaa would either need to be ionized and fully covering the source with a column of $N_{\rm H} = 1.4^{+0.6}_{-0.5} \times10^{22}$~cm$^{-2}$, or neutral and patchy, covering about 90 per cent of the central X-ray source.
Then, in just one week, all this obscuring material would need to clear out to produce the X-ray brightening.
It is hard to see how the stellar debris could disappear in such a short time. 
Fast ionized winds that have been seen, for example, from ASASSN-14li \citep{MKM2015}, reaching outflow velocities of up to $0.2\,c$ \citep{KDR2018}, and are much better candidates for the obscuring medium at early times of the TDE.
There are also TDEs like XMMSL1 J0740-85 \citep{SRK2017}, where X-ray spectral changes can be attributed to a variable ionized X-ray absorption and, more generally, occultations of the X-ray source by obscuring clouds are not uncommon in AGN \citep[e.g.,][]{2005ApJ...623L..93R,2010A&A...517A..47M,2011MNRAS.410.1027R,2014MNRAS.437.1776M}.
A cloud with a column density of $N_{\rm H} \sim 10^{22}$~cm$^{-2}$ in Keplerian orbit at $\sim 10^2-10^3\,r_g$ around a SMBH with $M_{BH}=3 \times 10^6 M_{\odot}$ could clear out in one week, given a density of a few $10^{6}-10^7$ cm$^{-3}$.
Such a radial range is consistent with the launching regions of accretion disk winds in AGN \citep[e.g.,][]{2000ApJ...543..686P,2010A&A...516A..89R}.

Yet another scenario for the late-time X-ray brightening of TDEs was recently proposed by \citet{WJS2020}.
In their ``thinning disk'' model, the initial X-ray obscuration would not be due to a wind, but rather caused by a geometrically thick disk obscuring the inner X-ray-emitting accretion flow, which in the cases of OGLE16aaa, ASASSN-15oi, and AT2019azh, would have to be viewed from a large inclination angle.
The late-time X-ray brightening would correspond to the thinning of the disk, possibly due to the accretion rate dropping down to the Eddington limit, which allows a direct view onto the X-ray emitting flow.
This model would nicely explain why the X-ray luminosity matches the Eddington limit in OGLE16aaa after the brightening and why it is only during the subsequent stages that the optical/X-ray spectral energy distribution is consistent with a thermal disc emission.

\section{Summary and conclusions}

OGLE16aaa represents the third known example of a late-time X-ray brightening in a TDE.
The factor of 20 X-ray brightening occurs in a time span of less than a week without any signs in the simultaneous optical/UV data, causing the X-ray emitting thermal component to be more than 17 times more luminous than the optical one.
For the first time we could pin-point the delay between the onset of the optical TDE and the X-ray brightening precisely to 182$\pm$5 days.
The bolometric luminosity fo the X-ray emitting black body component reached $L_{\rm bb,~bol} = (3.8\pm0.6)\times 10^{44}~\textrm{erg}~\textrm{s}^{-1}$, which matches the Eddington limit for the $3\times10^{6}$ solar mass black hole of OGLE16aaa.
However, the lack of X-ray temperature evolution is not consistent with the expected behavior for an accretion disk reaching the ISCO.

The delayed X-ray brightening of OGLE16aaa can be interpreted in two ways.
The initial optical/UV TDE could be generated by shocks in the circularization of the disrupted star around the SMBH. 
Then six months later, the majority of this gas finally accretes onto the black hole, thus causing the rapid X-ray brightening. 
Another scenario that cannot be ruled out is that initially the X-ray emitting inner accretion flow could be hidden behind a modest $N_{\rm H} \sim 10^{22}$~cm$^{-2}$ column of gas that is generated either by a wind or outflow, a geometrically thick disk, or the stellar debris, which then has to rapidly clear out between June 9 and June 16, 2016.

\begin{acknowledgements}
We would like to thank the referee for comments that helped to improve the manuscript.
JJEK and GM acknowledges support from the Spanish MINECO grant ESP2017-86582-C4-1-R. 
MG acknowledges support by the ``Programa de Atracci\'on de Talento'' of the Comunidad de Madrid, grant number 2018-T1/TIC-11733. 
This research has been partially funded by the AEI Project No. MDM-2017-0737 Unidad de Excelencia ``Mar\'{i}a de Maeztu'' - Centro de Astrobiolog\'{i}a (INTA-CSIC).
We acknowledge the use of public data from the Swift data archive. This work was partly based on observations obtained with XMM-Newton, an ESA science mission with instruments and contributions directly funded by ESA Member States and NASA.
\end{acknowledgements}

\bibliographystyle{aa}

\begin{thebibliography}{77}
\expandafter\ifx\csname natexlab\endcsname\relax\def\natexlab#1{#1}\fi

\bibitem[{{Auchettl} {et~al.}(2017){Auchettl}, {Guillochon}, \&
  {Ramirez-Ruiz}}]{AGRR2017}
{Auchettl}, K., {Guillochon}, J., \& {Ramirez-Ruiz}, E. 2017, \apj, 838, 149

\bibitem[{{Beloborodov}(1998)}]{Beloborodov1998}
{Beloborodov}, A.~M. 1998, \mnras, 297, 739

\bibitem[{{Blanchard} {et~al.}(2017){Blanchard}, {Nicholl}, {Berger},
  {Guillochon}, {Margutti}, {Chornock}, {Alexand er}, {Leja}, \&
  {Drout}}]{BNB2017}
{Blanchard}, P.~K., {Nicholl}, M., {Berger}, E., {et~al.} 2017, \apj, 843, 106

\bibitem[{{Bloom} {et~al.}(2011){Bloom}, {Giannios}, {Metzger}, {Cenko},
  {Perley}, {Butler}, {Tanvir}, {Levan}, {O'Brien}, {Strubbe}, {De Colle},
  {Ramirez-Ruiz}, {Lee}, {Nayakshin}, {Quataert}, {King}, {Cucchiara},
  {Guillochon}, {Bower}, {Fruchter}, {Morgan}, \& {van der Horst}}]{BGM2011}
{Bloom}, J.~S., {Giannios}, D., {Metzger}, B.~D., {et~al.} 2011, Science, 333,
  203

\bibitem[{{Bonnerot} {et~al.}(2017){Bonnerot}, {Rossi}, \& {Lodato}}]{BRL2017}
{Bonnerot}, C., {Rossi}, E.~M., \& {Lodato}, G. 2017, \mnras, 464, 2816

\bibitem[{{Brandt} \& {Alexander}(2015)}]{BrandtAlexander2015}
{Brandt}, W.~N. \& {Alexander}, D.~M. 2015, \aapr, 23, 1

\bibitem[{{Brown} {et~al.}(2016){Brown}, {Breeveld}, {Roming}, \&
  {Siegel}}]{BBR2016}
{Brown}, P.~J., {Breeveld}, A., {Roming}, P. W.~A., \& {Siegel}, M. 2016, \aj,
  152, 102

\bibitem[{{Burrows} {et~al.}(2011){Burrows}, {Kennea}, {Ghisellini}, {Mangano},
  {Zhang}, {Page}, {Eracleous}, {Romano}, {Sakamoto}, {Falcone}, {Osborne},
  {Campana}, {Beardmore}, {Breeveld}, {Chester}, {Corbet}, {Covino},
  {Cummings}, {D'Avanzo}, {D'Elia}, {Esposito}, {Evans}, {Fugazza}, {Gelbord},
  {Hiroi}, {Holland}, {Huang}, {Im}, {Israel}, {Jeon}, {Jeon}, {Jun}, {Kawai},
  {Kim}, {Krimm}, {Marshall}, {P. M{\'e}sz{\'a}ros}, {Negoro}, {Omodei},
  {Park}, {Perkins}, {Sugizaki}, {Sung}, {Tagliaferri}, {Troja}, {Ueda},
  {Urata}, {Usui}, {Antonelli}, {Barthelmy}, {Cusumano}, {Giommi}, {Melandri},
  {Perri}, {Racusin}, {Sbarufatti}, {Siegel}, \& {Gehrels}}]{BKG2011}
{Burrows}, D.~N., {Kennea}, J.~A., {Ghisellini}, G., {et~al.} 2011, \nat, 476,
  421

\bibitem[{{Cardelli} {et~al.}(1989){Cardelli}, {Clayton}, \&
  {Mathis}}]{CCM1989}
{Cardelli}, J.~A., {Clayton}, G.~C., \& {Mathis}, J.~S. 1989, \apj, 345, 245

\bibitem[{{Dai} {et~al.}(2018){Dai}, {McKinney}, {Roth}, {Ramirez-Ruiz}, \&
  {Miller}}]{DMR2018}
{Dai}, L., {McKinney}, J.~C., {Roth}, N., {Ramirez-Ruiz}, E., \& {Miller},
  M.~C. 2018, \apjl, 859, L20

\bibitem[{{Davis} \& {El-Abd}(2019)}]{Davis2019}
{Davis}, S.~W. \& {El-Abd}, S. 2019, \apj, 874, 23

\bibitem[{{Done} {et~al.}(2012){Done}, {Davis}, {Jin}, {Blaes}, \&
  {Ward}}]{DDJ2012}
{Done}, C., {Davis}, S.~W., {Jin}, C., {Blaes}, O., \& {Ward}, M. 2012, \mnras,
  420, 1848

\bibitem[{{Dong} {et~al.}(2016){Dong}, {Shappee}, {Prieto}, {Jha}, {Stanek},
  {Holoien}, {Kochanek}, {Thompson}, {Morrell}, {Thompson}, {Basu}, {Beacom},
  {Bersier}, {Brimacombe}, {Brown}, {Bufano}, {Chen}, {Conseil}, {Danilet},
  {Falco}, {Grupe}, {Kiyota}, {Masi}, {Nicholls}, {Olivares E.}, {Pignata},
  {Pojmanski}, {Simonian}, {Szczygiel}, \& {Wo{\'z}niak}}]{DSP2016}
{Dong}, S., {Shappee}, B.~J., {Prieto}, J.~L., {et~al.} 2016, Science, 351, 257

\bibitem[{{Esquej} {et~al.}(2007){Esquej}, {Saxton}, {Freyberg}, {Read},
  {Altieri}, {Sanchez-Portal}, \& {Hasinger}}]{ESF2007}
{Esquej}, P., {Saxton}, R.~D., {Freyberg}, M.~J., {et~al.} 2007, \aap, 462, L49

\bibitem[{{Evans} \& {Kochanek}(1989)}]{EvansKochanek1989}
{Evans}, C.~R. \& {Kochanek}, C.~S. 1989, \apjl, 346, L13

\bibitem[{{Evans} {et~al.}(2009){Evans}, {Beardmore}, {Page}, {Osborne},
  {O'Brien}, {Willingale}, {Starling}, {Burrows}, {Godet}, {Vetere}, {Racusin},
  {Goad}, {Wiersema}, {Angelini}, {Capalbi}, {Chincarini}, {Gehrels}, {Kennea},
  {Margutti}, {Morris}, {Mountford}, {Pagani}, {Perri}, {Romano}, \&
  {Tanvir}}]{EBP2009}
{Evans}, P.~A., {Beardmore}, A.~P., {Page}, K.~L., {et~al.} 2009, \mnras, 397,
  1177

\bibitem[{{Firth} {et~al.}(2016){Firth}, {Frohmaier}, {Dimitriadis}, {De Cia},
  {Galbany}, {Inserra}, {Kankare}, {Maguire}, {Smartt}, {Smith}, {Sullivan},
  {Valenti}, {Yaron}, {Young}, {Manulis}, {Baltay}, {Ellman}, {Hadjiyska},
  {McKinnon}, {Rabinowitz}, {Rostami}, {Feindt}, {Kowalski}, {Nugent}, \&
  {Wyrzykowski}}]{Firth2016ATel}
{Firth}, R., {Frohmaier}, C., {Dimitriadis}, G., {et~al.} 2016, The
  Astronomer's Telegram, 8559, 1

\bibitem[{{Gehrels} {et~al.}(2004){Gehrels}, {Chincarini}, {Giommi}, {Mason},
  {Nousek}, {Wells}, {White}, {Barthelmy}, {Burrows}, {Cominsky}, {Hurley},
  {Marshall}, {M{\'e}sz{\'a}ros}, {Roming}, {Angelini}, {Barbier}, {Belloni},
  {Campana}, {Caraveo}, {Chester}, {Citterio}, {Cline}, {Cropper}, {Cummings},
  {Dean}, {Feigelson}, {Fenimore}, {Frail}, {Fruchter}, {Garmire}, {Gendreau},
  {Ghisellini}, {Greiner}, {Hill}, {Hunsberger}, {Krimm}, {Kulkarni}, {Kumar},
  {Lebrun}, {Lloyd-Ronning}, {Markwardt}, {Mattson}, {Mushotzky}, {Norris},
  {Osborne}, {Paczynski}, {Palmer}, {Park}, {Parsons}, {Paul}, {Rees},
  {Reynolds}, {Rhoads}, {Sasseen}, {Schaefer}, {Short}, {Smale}, {Smith},
  {Stella}, {Tagliaferri}, {Takahashi}, {Tashiro}, {Townsley}, {Tueller},
  {Turner}, {Vietri}, {Voges}, {Ward}, {Willingale}, {Zerbi}, \&
  {Zhang}}]{Gehrels2004}
{Gehrels}, N., {Chincarini}, G., {Giommi}, P., {et~al.} 2004, \apj, 611, 1005

\bibitem[{{Gezari} {et~al.}(2017){Gezari}, {Cenko}, \& {Arcavi}}]{GCA2017}
{Gezari}, S., {Cenko}, S.~B., \& {Arcavi}, I. 2017, \apjl, 851, L47

\bibitem[{{Gierli{\'n}ski} \& {Done}(2004)}]{GD2004}
{Gierli{\'n}ski}, M. \& {Done}, C. 2004, \mnras, 349, L7

\bibitem[{{Giustini} {et~al.}(2020){Giustini}, {Miniutti}, \&
  {Saxton}}]{GMS2020}
{Giustini}, M., {Miniutti}, G., \& {Saxton}, R.~D. 2020, \aap, 636, L2

\bibitem[{{Guillochon} {et~al.}(2018){Guillochon}, {Nicholl}, {Villar},
  {Mockler}, {Narayan}, {Mandel}, {Berger}, \& {Williams}}]{GNV2018}
{Guillochon}, J., {Nicholl}, M., {Villar}, V.~A., {et~al.} 2018, \apjs, 236, 6

\bibitem[{{Guillochon} \& {Ramirez-Ruiz}(2013)}]{GRR2013}
{Guillochon}, J. \& {Ramirez-Ruiz}, E. 2013, \apj, 767, 25

\bibitem[{{G{\"u}ver} \& {{\"O}zel}(2009)}]{GO2009}
{G{\"u}ver}, T. \& {{\"O}zel}, F. 2009, \mnras, 400, 2050

\bibitem[{{Heckman} \& {Best}(2014)}]{HeckmanBest2014}
{Heckman}, T.~M. \& {Best}, P.~N. 2014, \araa, 52, 589

\bibitem[{{HI4PI Collaboration} {et~al.}(2016){HI4PI Collaboration}, {Ben
  Bekhti}, {Fl{\"o}er}, {Keller}, {Kerp}, {Lenz}, {Winkel}, {Bailin},
  {Calabretta}, {Dedes}, {Ford}, {Gibson}, {Haud}, {Janowiecki}, {Kalberla},
  {Lockman}, {McClure-Griffiths}, {Murphy}, {Nakanishi}, {Pisano}, \&
  {Staveley-Smith}}]{HI4PI2016}
{HI4PI Collaboration}, {Ben Bekhti}, N., {Fl{\"o}er}, L., {et~al.} 2016, \aap,
  594, A116

\bibitem[{{Holoien} {et~al.}(2018){Holoien}, {Brown}, {Auchettl}, {Kochanek},
  {Prieto}, {Shappee}, \& {Van Saders}}]{HBA2018}
{Holoien}, T.~W.~S., {Brown}, J.~S., {Auchettl}, K., {et~al.} 2018, \mnras,
  480, 5689

\bibitem[{{Holoien} {et~al.}(2016){Holoien}, {Kochanek}, {Prieto}, {Stanek},
  {Dong}, {Shappee}, {Grupe}, {Brown}, {Basu}, {Beacom}, {Bersier},
  {Brimacombe}, {Danilet}, {Falco}, {Guo}, {Jose}, {Herczeg}, {Long},
  {Pojmanski}, {Simonian}, {Szczygie{\l}}, {Thompson}, {Thorstensen}, {Wagner},
  \& {Wo{\'z}niak}}]{HKP2016}
{Holoien}, T.~W.~S., {Kochanek}, C.~S., {Prieto}, J.~L., {et~al.} 2016, \mnras,
  455, 2918

\bibitem[{{Jiang} {et~al.}(2019){Jiang}, {Wang}, {Mou}, {Liu}, {Dou}, {Sheng},
  \& {Wang}}]{JNW2019}
{Jiang}, N., {Wang}, T., {Mou}, G., {et~al.} 2019, \apj, 871, 15

\bibitem[{{Jiang} {et~al.}(2017){Jiang}, {Wang}, {Yan}, {Xiao}, {Yang}, {Dou},
  {Wang}, {Cutri}, \& {Mainzer}}]{JWY2017}
{Jiang}, N., {Wang}, T., {Yan}, L., {et~al.} 2017, \apj, 850, 63

\bibitem[{{Kankare} {et~al.}(2017){Kankare}, {Kotak}, {Mattila}, {Lundqvist},
  {Ward}, {Fraser}, {Lawrence}, {Smartt}, {Meikle}, {Bruce}, {Harmanen},
  {Hutton}, {Inserra}, {Kangas}, {Pastorello}, {Reynolds},
  {Romero-Ca{\~n}izales}, {Smith}, {Valenti}, {Chambers}, {Hodapp}, {Huber},
  {Kaiser}, {Kudritzki}, {Magnier}, {Tonry}, {Wainscoat}, \&
  {Waters}}]{KKM2017}
{Kankare}, E., {Kotak}, R., {Mattila}, S., {et~al.} 2017, Nature Astronomy, 1,
  865

\bibitem[{{Kara} {et~al.}(2018){Kara}, {Dai}, {Reynolds}, \&
  {Kallman}}]{KDR2018}
{Kara}, E., {Dai}, L., {Reynolds}, C.~S., \& {Kallman}, T. 2018, \mnras, 474,
  3593

\bibitem[{{Kormendy} \& {Ho}(2013)}]{KormendyHo2013}
{Kormendy}, J. \& {Ho}, L.~C. 2013, \araa, 51, 511

\bibitem[{{Krolik} {et~al.}(2016){Krolik}, {Piran}, {Svirski}, \&
  {Cheng}}]{KPS2016}
{Krolik}, J., {Piran}, T., {Svirski}, G., \& {Cheng}, R.~M. 2016, \apj, 827,
  127

\bibitem[{{Lacy} {et~al.}(1982){Lacy}, {Townes}, \& {Hollenbach}}]{LTH1982}
{Lacy}, J.~H., {Townes}, C.~H., \& {Hollenbach}, D.~J. 1982, \apj, 262, 120

\bibitem[{{Leloudas} {et~al.}(2016){Leloudas}, {Fraser}, {Stone}, {van Velzen},
  {Jonker}, {Arcavi}, {Fremling}, {Maund}, {Smartt}, {Kr{\`\i}hler},
  {Miller-Jones}, {Vreeswijk}, {Gal-Yam}, {Mazzali}, {De Cia}, {Howell},
  {Inserra}, {Patat}, {de Ugarte Postigo}, {Yaron}, {Ashall}, {Bar},
  {Campbell}, {Chen}, {Childress}, {Elias-Rosa}, {Harmanen}, {Hosseinzadeh},
  {Johansson}, {Kangas}, {Kankare}, {Kim}, {Kuncarayakti}, {Lyman}, {Magee},
  {Maguire}, {Malesani}, {Mattila}, {McCully}, {Nicholl}, {Prentice},
  {Romero-Ca{\~n}izales}, {Schulze}, {Smith}, {Sollerman}, {Sullivan},
  {Tucker}, {Valenti}, {Wheeler}, \& {Young}}]{LFS2016}
{Leloudas}, G., {Fraser}, M., {Stone}, N.~C., {et~al.} 2016, Nature Astronomy,
  1, 0002

\bibitem[{{Lin} {et~al.}(2015){Lin}, {Maksym}, {Irwin}, {Komossa}, {Webb},
  {Godet}, {Barret}, {Grupe}, \& {Gwyn}}]{LMI2015}
{Lin}, D., {Maksym}, P.~W., {Irwin}, J.~A., {et~al.} 2015, \apj, 811, 43

\bibitem[{{Liu} {et~al.}(2019){Liu}, {Dou}, {Shen}, \& {Chen}}]{LDS2019}
{Liu}, X.-L., {Dou}, L.-M., {Shen}, R.-F., \& {Chen}, J.-H. 2019,
  arXiv:1912.06081

\bibitem[{{Lodato} {et~al.}(2009){Lodato}, {King}, \& {Pringle}}]{LKP2009}
{Lodato}, G., {King}, A.~R., \& {Pringle}, J.~E. 2009, \mnras, 392, 332

\bibitem[{{Loeb} \& {Ulmer}(1997)}]{LU1997}
{Loeb}, A. \& {Ulmer}, A. 1997, \apj, 489, 573

\bibitem[{{Lu} \& {Bonnerot}(2020)}]{LuBonnerot2020}
{Lu}, W. \& {Bonnerot}, C. 2020, \mnras, 492, 686

\bibitem[{{Maiolino} {et~al.}(2010){Maiolino}, {Risaliti}, {Salvati},
  {Pietrini}, {Torricelli-Ciamponi}, {Elvis}, {Fabbiano}, {Braito}, \&
  {Reeves}}]{2010A&A...517A..47M}
{Maiolino}, R., {Risaliti}, G., {Salvati}, M., {et~al.} 2010, \aap, 517, A47

\bibitem[{{Metzger} \& {Stone}(2016)}]{MS2016}
{Metzger}, B.~D. \& {Stone}, N.~C. 2016, \mnras, 461, 948

\bibitem[{{Miller} {et~al.}(2015){Miller}, {Kaastra}, {Miller}, {Reynolds},
  {Brown}, {Cenko}, {Drake}, {Gezari}, {Guillochon}, {Gultekin}, {Irwin},
  {Levan}, {Maitra}, {Maksym}, {Mushotzky}, {O'Brien}, {Paerels}, {de Plaa},
  {Ramirez-Ruiz}, {Strohmayer}, \& {Tanvir}}]{MKM2015}
{Miller}, J.~M., {Kaastra}, J.~S., {Miller}, M.~C., {et~al.} 2015, \nat, 526,
  542

\bibitem[{{Miniutti} {et~al.}(2014){Miniutti}, {Sanfrutos}, {Beuchert},
  {Ag{\'\i}s-Gonz{\'a}lez}, {Longinotti}, {Piconcelli}, {Krongold},
  {Guainazzi}, {Bianchi}, {Matt}, \&
  {Jim{\'e}nez-Bail{\'o}n}}]{2014MNRAS.437.1776M}
{Miniutti}, G., {Sanfrutos}, M., {Beuchert}, T., {et~al.} 2014, \mnras, 437,
  1776

\bibitem[{{Miniutti} {et~al.}(2019){Miniutti}, {Saxton}, {Giustini}, {Alexand
  er}, {Fender}, {Heywood}, {Monageng}, {Coriat}, {Tzioumis}, {Read}, {Knigge},
  {Gandhi}, {Pretorius}, \& {Ag{\'\i}s-Gonz{\'a}lez}}]{MSG2019}
{Miniutti}, G., {Saxton}, R.~D., {Giustini}, M., {et~al.} 2019, \nat, 573, 381

\bibitem[{{Mitsuda} {et~al.}(1984){Mitsuda}, {Inoue}, {Koyama}, {Makishima},
  {Matsuoka}, {Ogawara}, {Shibazaki}, {Suzuki}, {Tanaka}, \&
  {Hirano}}]{MIK1984}
{Mitsuda}, K., {Inoue}, H., {Koyama}, K., {et~al.} 1984, \pasj, 36, 741

\bibitem[{{Mockler} {et~al.}(2019){Mockler}, {Guillochon}, \&
  {Ramirez-Ruiz}}]{MGRR2019}
{Mockler}, B., {Guillochon}, J., \& {Ramirez-Ruiz}, E. 2019, \apj, 872, 151

\bibitem[{{Mummery} \& {Balbus}(2020)}]{MummeryBalbus2020}
{Mummery}, A. \& {Balbus}, S.~A. 2020, \mnras, 492, 5655

\bibitem[{{Neustadt} {et~al.}(2020){Neustadt}, {Holoien}, {Kochanek},
  {Auchettl}, {Brown}, {Shappee}, {Pogge}, {Dong}, {Stanek}, {Tucker}, {Bose},
  {Chen}, {Ricci}, {Vallely}, {Prieto}, {Thompson}, {Coulter}, {Drout},
  {Foley}, {Kilpatrick}, {Piro}, {Rojas-Bravo}, {Buckley}, {Gromadzki},
  {Dimitriadis}, {Siebert}, {Do}, {Huber}, \& {Payne}}]{NHK2020}
{Neustadt}, J.~M.~M., {Holoien}, T.~W.~S., {Kochanek}, C.~S., {et~al.} 2020,
  \mnras, 494, 2538

\bibitem[{{Onori} {et~al.}(2019){Onori}, {Cannizzaro}, {Jonker}, {Fraser},
  {Kostrzewa-Rutkowska}, {Martin-Carrillo}, {Benetti}, {Elias-Rosa},
  {Gromadzki}, {Harmanen}, {Mattila}, {Strizinger}, {Terreran}, \&
  {Wevers}}]{OCJ2019}
{Onori}, F., {Cannizzaro}, G., {Jonker}, P.~G., {et~al.} 2019, \mnras, 489,
  1463

\bibitem[{{Phinney}(1989)}]{Phinney1989}
{Phinney}, E.~S. 1989, in IAU Symposium, Vol. 136, The Center of the Galaxy,
  ed. M.~{Morris}, 543

\bibitem[{{Piran} {et~al.}(2015){Piran}, {Svirski}, {Krolik}, {Cheng}, \&
  {Shiokawa}}]{PSK2015}
{Piran}, T., {Svirski}, G., {Krolik}, J., {Cheng}, R.~M., \& {Shiokawa}, H.
  2015, \apj, 806, 164

\bibitem[{{Planck Collaboration} {et~al.}(2016){Planck Collaboration}, {Ade},
  {Aghanim}, {Arnaud}, {Ashdown}, {Aumont}, {Baccigalupi}, {Banday},
  {Barreiro}, {Bartlett}, {Bartolo}, {Battaner}, {Battye}, {Benabed},
  {Beno{\^\i}t}, {Benoit-L{\'e}vy}, {Bernard}, {Bersanelli}, {Bielewicz},
  {Bock}, {Bonaldi}, {Bonavera}, {Bond}, {Borrill}, {Bouchet}, {Boulanger},
  {Bucher}, {Burigana}, {Butler}, {Calabrese}, {Cardoso}, {Catalano},
  {Challinor}, {Chamballu}, {Chary}, {Chiang}, {Chluba}, {Christensen},
  {Church}, {Clements}, {Colombi}, {Colombo}, {Combet}, {Coulais}, {Crill},
  {Curto}, {Cuttaia}, {Danese}, {Davies}, {Davis}, {de Bernardis}, {de Rosa},
  {de Zotti}, {Delabrouille}, {D{\'e}sert}, {Di Valentino}, {Dickinson},
  {Diego}, {Dolag}, {Dole}, {Donzelli}, {Dor{\'e}}, {Douspis}, {Ducout},
  {Dunkley}, {Dupac}, {Efstathiou}, {Elsner}, {En{\ss}lin}, {Eriksen},
  {Farhang}, {Fergusson}, {Finelli}, {Forni}, {Frailis}, {Fraisse},
  {Franceschi}, {Frejsel}, {Galeotta}, {Galli}, {Ganga}, {Gauthier}, {Gerbino},
  {Ghosh}, {Giard}, {Giraud-H{\'e}raud}, {Giusarma}, {Gjerl{\o}w},
  {Gonz{\'a}lez-Nuevo}, {G{\'o}rski}, {Gratton}, {Gregorio}, {Gruppuso},
  {Gudmundsson}, {Hamann}, {Hansen}, {Hanson}, {Harrison}, {Helou},
  {Henrot-Versill{\'e}}, {Hern{\'a}ndez-Monteagudo}, {Herranz}, {Hildebrand t},
  {Hivon}, {Hobson}, {Holmes}, {Hornstrup}, {Hovest}, {Huang}, {Huffenberger},
  {Hurier}, {Jaffe}, {Jaffe}, {Jones}, {Juvela}, {Keih{\"a}nen}, {Keskitalo},
  {Kisner}, {Kneissl}, {Knoche}, {Knox}, {Kunz}, {Kurki-Suonio}, {Lagache},
  {L{\"a}hteenm{\"a}ki}, {Lamarre}, {Lasenby}, {Lattanzi}, {Lawrence}, {Leahy},
  {Leonardi}, {Lesgourgues}, {Levrier}, {Lewis}, {Liguori}, {Lilje},
  {Linden-V{\o}rnle}, {L{\'o}pez-Caniego}, {Lubin}, {Mac{\'\i}as-P{\'e}rez},
  {Maggio}, {Maino}, {Mandolesi}, {Mangilli}, {Marchini}, {Maris}, {Martin},
  {Martinelli}, {Mart{\'\i}nez-Gonz{\'a}lez}, {Masi}, {Matarrese}, {McGehee},
  {Meinhold}, {Melchiorri}, {Melin}, {Mendes}, {Mennella}, {Migliaccio},
  {Millea}, {Mitra}, {Miville-Desch{\^e}nes}, {Moneti}, {Montier}, {Morgante},
  {Mortlock}, {Moss}, {Munshi}, {Murphy}, {Naselsky}, {Nati}, {Natoli},
  {Netterfield}, {N{\o}rgaard-Nielsen}, {Noviello}, {Novikov}, {Novikov},
  {Oxborrow}, {Paci}, {Pagano}, {Pajot}, {Paladini}, {Paoletti}, {Partridge},
  {Pasian}, {Patanchon}, {Pearson}, {Perdereau}, {Perotto}, {Perrotta},
  {Pettorino}, {Piacentini}, {Piat}, {Pierpaoli}, {Pietrobon}, {Plaszczynski},
  {Pointecouteau}, {Polenta}, {Popa}, {Pratt}, {Pr{\'e}zeau}, {Prunet},
  {Puget}, {Rachen}, {Reach}, {Rebolo}, {Reinecke}, {Remazeilles}, {Renault},
  {Renzi}, {Ristorcelli}, {Rocha}, {Rosset}, {Rossetti}, {Roudier},
  {Rouill{\'e} d'Orfeuil}, {Rowan-Robinson}, {Rubi{\~n}o-Mart{\'\i}n},
  {Rusholme}, {Said}, {Salvatelli}, {Salvati}, {Sandri}, {Santos},
  {Savelainen}, {Savini}, {Scott}, {Seiffert}, {Serra}, {Shellard}, {Spencer},
  {Spinelli}, {Stolyarov}, {Stompor}, {Sudiwala}, {Sunyaev}, {Sutton},
  {Suur-Uski}, {Sygnet}, {Tauber}, {Terenzi}, {Toffolatti}, {Tomasi},
  {Tristram}, {Trombetti}, {Tucci}, {Tuovinen}, {T{\"u}rler}, {Umana},
  {Valenziano}, {Valiviita}, {Van Tent}, {Vielva}, {Villa}, {Wade}, {Wandelt},
  {Wehus}, {White}, {White}, {Wilkinson}, {Yvon}, {Zacchei}, \&
  {Zonca}}]{Planck2016}
{Planck Collaboration}, {Ade}, P.~A.~R., {Aghanim}, N., {et~al.} 2016, \aap,
  594, A13

\bibitem[{{Poole} {et~al.}(2008){Poole}, {Breeveld}, {Page}, {Land sman},
  {Holland}, {Roming}, {Kuin}, {Brown}, {Gronwall}, {Hunsberger}, {Koch},
  {Mason}, {Schady}, {vanden Berk}, {Blustin}, {Boyd}, {Broos}, {Carter},
  {Chester}, {Cucchiara}, {Hancock}, {Huckle}, {Immler}, {Ivanushkina},
  {Kennedy}, {Marshall}, {Morgan}, {Pandey}, {de Pasquale}, {Smith}, \&
  {Still}}]{Poole2008}
{Poole}, T.~S., {Breeveld}, A.~A., {Page}, M.~J., {et~al.} 2008, \mnras, 383,
  627

\bibitem[{{Pringle}(1981)}]{Pringle1981}
{Pringle}, J.~E. 1981, \araa, 19, 137

\bibitem[{{Proga} {et~al.}(2000){Proga}, {Stone}, \&
  {Kallman}}]{2000ApJ...543..686P}
{Proga}, D., {Stone}, J.~M., \& {Kallman}, T.~R. 2000, \apj, 543, 686

\bibitem[{{Rees}(1988)}]{Rees1988}
{Rees}, M.~J. 1988, \nat, 333, 523

\bibitem[{{Rees}(1990)}]{Rees1990}
{Rees}, M.~J. 1990, Science, 247, 817

\bibitem[{{Risaliti} \& {Elvis}(2010)}]{2010A&A...516A..89R}
{Risaliti}, G. \& {Elvis}, M. 2010, \aap, 516, A89

\bibitem[{{Risaliti} {et~al.}(2005){Risaliti}, {Elvis}, {Fabbiano}, {Baldi}, \&
  {Zezas}}]{2005ApJ...623L..93R}
{Risaliti}, G., {Elvis}, M., {Fabbiano}, G., {Baldi}, A., \& {Zezas}, A. 2005,
  \apjl, 623, L93

\bibitem[{{Risaliti} {et~al.}(2011){Risaliti}, {Nardini}, {Salvati}, {Elvis},
  {Fabbiano}, {Maiolino}, {Pietrini}, \&
  {Torricelli-Ciamponi}}]{2011MNRAS.410.1027R}
{Risaliti}, G., {Nardini}, E., {Salvati}, M., {et~al.} 2011, \mnras, 410, 1027

\bibitem[{{Saxton} {et~al.}(2017){Saxton}, {Read}, {Komossa}, {Lira},
  {Alexander}, \& {Wieringa}}]{SRK2017}
{Saxton}, R.~D., {Read}, A.~M., {Komossa}, S., {et~al.} 2017, \aap, 598, A29

\bibitem[{{Schlafly} \& {Finkbeiner}(2011)}]{SF2011}
{Schlafly}, E.~F. \& {Finkbeiner}, D.~P. 2011, \apj, 737, 103

\bibitem[{{Shiokawa} {et~al.}(2015){Shiokawa}, {Krolik}, {Cheng}, {Piran}, \&
  {Noble}}]{SKC2015}
{Shiokawa}, H., {Krolik}, J.~H., {Cheng}, R.~M., {Piran}, T., \& {Noble}, S.~C.
  2015, \apj, 804, 85

\bibitem[{{Strubbe} \& {Quataert}(2009)}]{StrubbeQuataert2009}
{Strubbe}, L.~E. \& {Quataert}, E. 2009, \mnras, 400, 2070

\bibitem[{{Tchekhovskoy} {et~al.}(2014){Tchekhovskoy}, {Metzger}, {Giannios},
  \& {Kelley}}]{TMG2014}
{Tchekhovskoy}, A., {Metzger}, B.~D., {Giannios}, D., \& {Kelley}, L.~Z. 2014,
  \mnras, 437, 2744

\bibitem[{{Trakhtenbrot} {et~al.}(2019){Trakhtenbrot}, {Arcavi}, {MacLeod},
  {Ricci}, {Kara}, {Graham}, {Stern}, {Harrison}, {Burke}, {Hiramatsu},
  {Hosseinzadeh}, {Howell}, {Smartt}, {Rest}, {Prieto}, {Shappee}, {Holoien},
  {Bersier}, {Filippenko}, {Brink}, {Zheng}, {Li}, {Remillard}, \&
  {Loewenstein}}]{TAM2019}
{Trakhtenbrot}, B., {Arcavi}, I., {MacLeod}, C.~L., {et~al.} 2019, \apj, 883,
  94

\bibitem[{{van Velzen} {et~al.}(2016){van Velzen}, {Anderson}, {Stone},
  {Fraser}, {Wevers}, {Metzger}, {Jonker}, {van der Horst}, {Staley}, {Mendez},
  {Miller-Jones}, {Hodgkin}, {Campbell}, \& {Fender}}]{vVAS2016}
{van Velzen}, S., {Anderson}, G.~E., {Stone}, N.~C., {et~al.} 2016, Science,
  351, 62

\bibitem[{{van Velzen} {et~al.}(2020){van Velzen}, {Gezari}, {Hammerstein},
  {Roth}, {Frederick}, {Ward}, {Hung}, {Cenko}, {Stein}, {Perley}, {Taggart},
  {Sollerman}, {Andreoni}, {Bellm}, {Brinnel}, {De}, {Dekany}, {Feeney},
  {Foley}, {Fremling}, {Giomi}, {Golkhou}, {Ho}, {Kasliwal}, {Kilpatrick},
  {Kulkarni}, {Kupfer}, {Laher}, {Mahabal}, {Masci}, {Nordin}, {Riddle},
  {Rusholme}, {Sharma}, {van Santen}, {Shupe}, \& {Soumagnac}}]{vVGH2020}
{van Velzen}, S., {Gezari}, S., {Hammerstein}, E., {et~al.} 2020,
  arXiv:2001.01409

\bibitem[{{Wen} {et~al.}(2020){Wen}, {Jonker}, {Stone}, {Zabludoff}, \&
  {Psaltis}}]{WJS2020}
{Wen}, S., {Jonker}, P.~G., {Stone}, N.~C., {Zabludoff}, A.~I., \& {Psaltis},
  D. 2020, arXiv:2003.12583

\bibitem[{{Wevers} {et~al.}(2019){Wevers}, {Pasham}, {van Velzen}, {Leloudas},
  {Schulze}, {Miller-Jones}, {Jonker}, {Gromadzki}, {Kankare}, {Hodgkin},
  {Wyrzykowski}, {Kostrzewa-Rutkowska}, {Moran}, {Berton}, {Maguire}, {Onori},
  {Mattila}, \& {Nicholl}}]{WPvV2019}
{Wevers}, T., {Pasham}, D.~R., {van Velzen}, S., {et~al.} 2019, \mnras, 488,
  4816

\bibitem[{{Wilms} {et~al.}(2000){Wilms}, {Allen}, \& {McCray}}]{WAM2000}
{Wilms}, J., {Allen}, A., \& {McCray}, R. 2000, \apj, 542, 914

\bibitem[{{Wyrzykowski} {et~al.}(2014){Wyrzykowski}, {Kostrzewa-Rutkowska},
  {Koz{\l}owski}, {Udalski}, {Poleski}, {Skowron}, {Blagorodnova}, {Kubiak},
  {Szyma{\'n}ski}, {Pietrzy{\'n}ski}, {Soszy{\'n}ski}, {Ulaczyk},
  {Pietrukowicz}, \& {Mr{\'o}z}}]{Wyrzykowski2014}
{Wyrzykowski}, {\L}., {Kostrzewa-Rutkowska}, Z., {Koz{\l}owski}, S., {et~al.}
  2014, \actaa, 64, 197

\bibitem[{{Wyrzykowski} {et~al.}(2016){Wyrzykowski}, {Kostrzewa-Rutkowska},
  {Udalski}, {Kozlowski}, {Pawlak}, {Szymanski}, {Sitek}, {Klencki}, \&
  {Arcavi}}]{WKU2016ATel}
{Wyrzykowski}, L., {Kostrzewa-Rutkowska}, Z., {Udalski}, A., {et~al.} 2016, The
  Astronomer's Telegram, 8577, 1

\bibitem[{{Wyrzykowski} {et~al.}(2017){Wyrzykowski}, {Zieli{\'n}ski},
  {Kostrzewa-Rutkowska}, {Hamanowicz}, {Jonker}, {Arcavi}, {Guillochon},
  {Brown}, {Koz{\l}owski}, {Udalski}, {Szyma{\'n}ski}, {Soszy{\'n}ski},
  {Poleski}, {Pietrukowicz}, {Skowron}, {Mr{\'o}z}, {Ulaczyk}, {Pawlak},
  {Rybicki}, {Greiner}, {Kr{\"u}hler}, {Bolmer}, {Smartt}, {Maguire}, \&
  {Smith}}]{WLZ2017}
{Wyrzykowski}, {\L}., {Zieli{\'n}ski}, M., {Kostrzewa-Rutkowska}, Z., {et~al.}
  2017, \mnras, 465, L114

\bibitem[{{Yang} {et~al.}(2013){Yang}, {Wang}, {Ferland}, {Yuan}, {Zhou}, \&
  {Jiang}}]{YCW2013}
{Yang}, C.-W., {Wang}, T.-G., {Ferland}, G., {et~al.} 2013, \apj, 774, 46

\end{thebibliography}


\end{document}